\documentclass[prb,superscriptaddress,twocolumn,showpacs,preprintnumbers,amsmath,amssymb]{revtex4} 

\usepackage{graphicx} 
\usepackage{dcolumn} % Align table columns on decimal point
\usepackage{bm}
\usepackage{txfonts}
\usepackage{mathrsfs}
\usepackage{comment}

\usepackage[]{psfrag}

\psfrag{tyldu}{$u_0$}
\psfrag{deltal}{$\delta$}
\psfrag{logdelt}{$\log |a_2-a_2^{(0)}|$}
\psfrag{logphi}{$\log (\phi_0)$}
\psfrag{dgdc}{$\frac{\delta_G-\delta_c}{\delta_c}$}
\psfrag{Temp}{$T$}
\psfrag{etaom}{$\eta_\omega$}
\psfrag{aa22}{$a_2$}
\psfrag{aa44}{$a_4$}
\psfrag{rown1}{$a_4^2=3a_2a_6$}
\psfrag{rown2}{$a_4^2=4a_2a_6$}
\psfrag{logTTT}{$\log(T^{tri}-T)$}
\psfrag{rhobar}{\hspace{0.5cm}$\bar{\rho}$}
\psfrag{log(tyldeaaa)}{$log|a^0_2-a_2^{0,\rm cr}|$}

\begin{document}

%\preprint{ }

\title{Finite temperature crossovers near quantum tricritical points in metals}
%%%\title{Finite temperature crossovers near metallic quantum tricritical points}

\author{P. Jakubczyk}
\email{Pawel.Jakubczyk@fuw.edu.pl}
\affiliation{Institute for Theoretical Physics, Warsaw University, Ho\.za 69, 
 00-681 Warsaw, Poland}
\affiliation{Max-Planck-Institute for Solid State Research, Heisenbergstr.~1, 
 D-70569 Stuttgart, Germany } 
\author{J. Bauer}
\affiliation{Max-Planck-Institute for Solid State Research, Heisenbergstr.~1, 
 D-70569 Stuttgart, Germany } 
\author{W. Metzner}
\affiliation{Max-Planck-Institute for Solid State Research,
Heisenbergstr.~1, D-70569 Stuttgart, Germany }

\date{\today}

\begin{abstract}
We present a renormalization group treatment of quantum tricriticality in
metals. Applying a set of flow equations derived within the functional
renormalization group framework we evaluate the correlation length in the quantum 
critical region of the phase diagram, extending into finite temperatures above the 
quantum critical or tricritical point. We calculate the finite temperature phase
boundaries and analyze the crossover behavior when the system is tuned between
quantum criticality and quantum tricriticality.
%%%quantum critical and tricritical scenarios.   
\end{abstract}

\maketitle

\section{Introduction}

Quantum critical behavior in metals is a topic of prime interest for the
theory and experiment in the field of condensed matter physics.
\cite{sachdev_book, vojta_review03, loehneysen_review06, belitz_review05, 
abanov03, gegenwart08,stewart01} Despite significant 
effort made over the last couple of years, several intriguing puzzles remain 
unresolved by the quickly developing theory of quantum phase transitions. 
These include both, fundamental problems regarding the correct low energy action 
to describe quantum critical systems, and the physical mechanisms responsible for 
unusual behavior of specific compounds. 

In the conventional scenario of quantum criticality the critical temperature
$T_c$ for a second order transition is driven to zero by a tuning parameter,
so that a quantum critical point emerges. However, it is also possible that
below a tricritical temperature the transition becomes first
order. Then a particular brand of quantum criticality
arises when the finite temperature ($T$) phase boundary terminates at $T=0$
with the tricritical point. Strictly speaking, in real systems this scenario is 
hardly feasible, since it requires fine tuning of two non-thermal parameters. 
However, proximity to quantum tricriticality has been invoked to explain a number 
of experimentally observed properties of metallic compounds.
\cite{pfleiderer01, uhlarz04, belitz05, green05, misawa08, misawa09} 
%%%For example, singular behavior of magnetic susceptibilities at both $q=0$ and 
%%%the ordering wavevector $q=Q$ offers a route to explain low-temperature properties 
%%%of a number of antiferromagnetic heavy-fermion compounds,\cite{misawa08, misawa09} 
%%%posing an alternative to earlier theory which invoked Kondo physics.\cite{gegenwart08}

Although the relevance of quantum tricriticality was emphasized in a number of recent 
works,\cite{belitz05, green05, misawa08, misawa09} a renormalization group (RG) study 
addressing crossovers between critical and tricritical behavior, in particular in the 
quantum-critical, non-Fermi liquid regime, is missing. 
%In this work we aim at filling this gap. 
In this paper we present a renormalization group analysis for a model system 
displaying such crossovers.
We build upon an earlier work,\cite{jakubczyk09} where renormalization group 
flows of a quantum $\phi^6$-model were already studied. The main focus of that
work was to show that order parameter fluctuations may turn first order transitions
occurring within the bare model into continuous transitions.   

The paper is structured as follows. In Sec.~II we present a phenomenological
picture of the problem to be analyzed, and in Sec.~III we define the corresponding
model action. 
%%%In Sec. III we introduce the model action applied in our following RG procedure. 
In Sec.~IV we describe the functional RG framework and derive the RG flow equations. 
In Sec.~V we analyze the linearized forms of the flow equations obtaining
analytical solutions in spatial dimensions $d>2$. 
In Sec.~VI we present example numerical results for the 
correlation length in the quantum critical regime, and the shapes of the phase
boundaries. In the final Sec.~VII we lay out our conclusions.      

%%%%%%%%%%%%%%%%%%%%%%%%%%%%%%%%%%%%%%%%%%%%%%%%%%%%%%%%%%%%%%%%%%%%%%%%%%%%%%%%%%%%%%%%%%

\section{Phenomenological picture}

\begin{figure}[ht!]
\begin{center}
\includegraphics[width=2.1in]{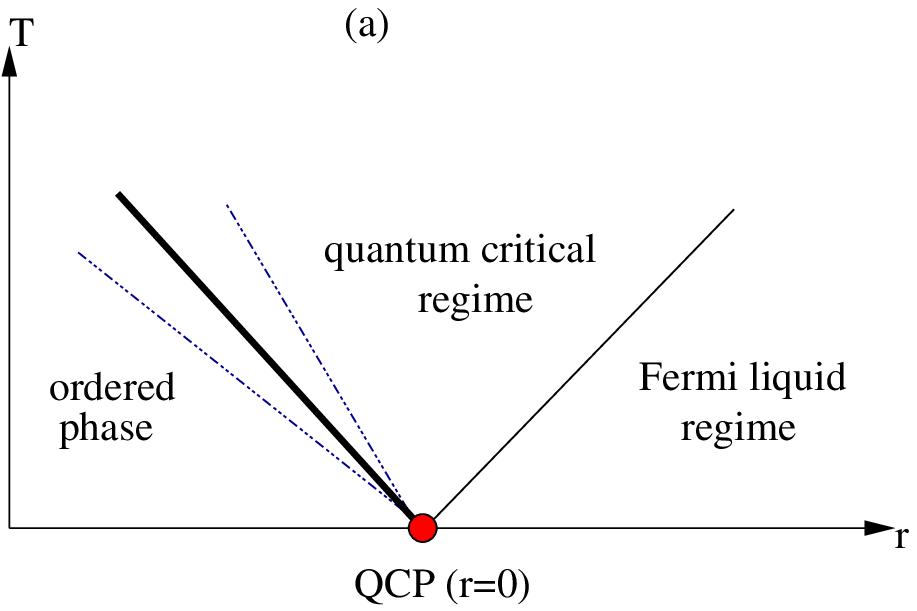}
\includegraphics[width=2.1in]{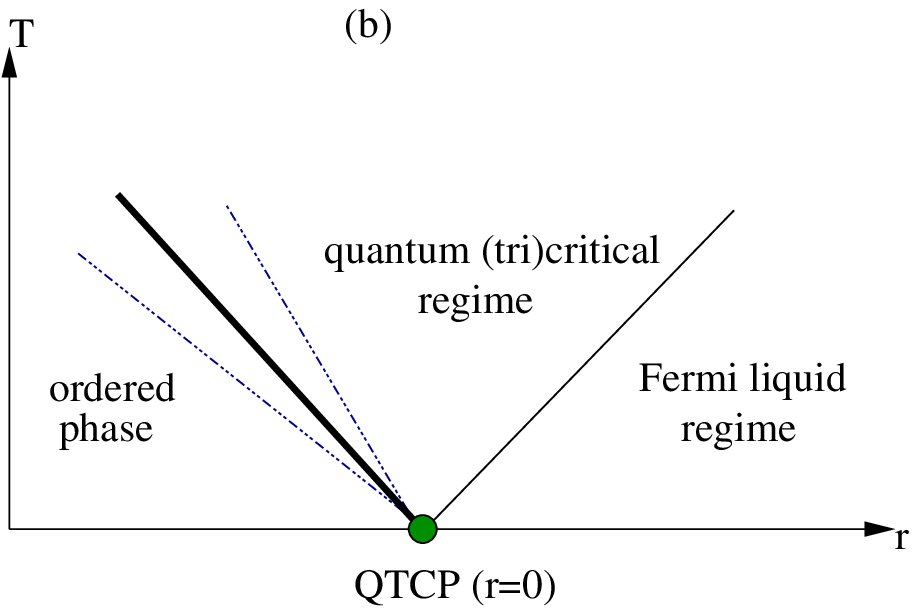}
\includegraphics[width=2.1in]{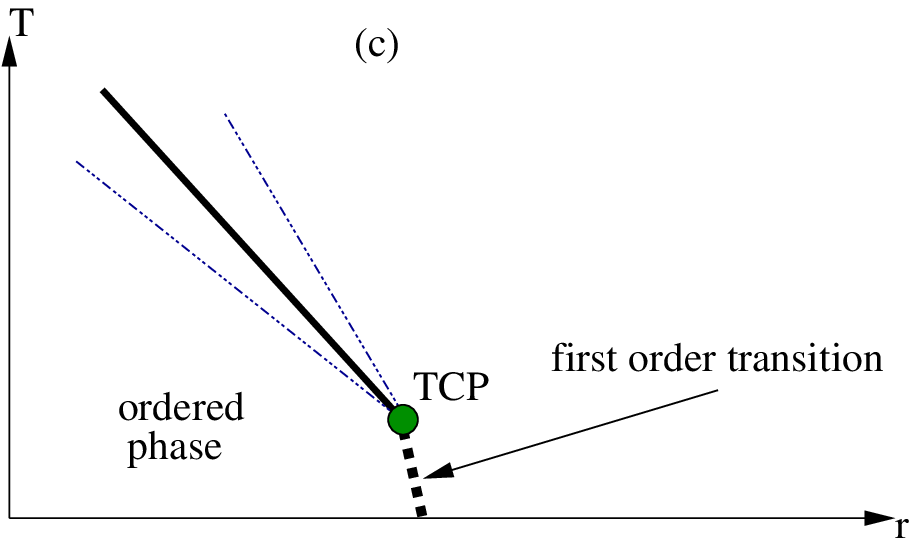}
\caption{(Color online) Different scenarios for generic phase diagrams. 
  Bold solid (dotted) lines mark second (first) order phase transitions,
  thin solid lines indicate the crossover between quantum critical and
  Fermi liquid behavior, while the thin dotted lines indicate the 
  boundaries of the Ginzburg region.
  (a) Schematic phase diagram of a system exhibiting a 
  quantum critical point, at which a line of second order finite $T$ phase
  transitions terminates. A crossover region, represented here with a line, separates 
  the Fermi liquid and quantum critical regimes in the disordered phase. 
  (b) Schematic phase diagram of a system exhibiting a quantum tricritical point.
  (c) Schematic phase diagram of a system exhibiting a tricritical point at a finite
  temperature. The transition is first order below the tricritical
  temperature. Critical quantum fluctuations are absent. Varying another
  control coupling one can tune between the depicted situations (see the main
  text).} 
\label{Phase_diag_pheno}
\end{center}
\end{figure}

Before plunging into the details of the formalism let us consider the
different scenarios which are envisaged in the following.
In Fig.~\ref{Phase_diag_pheno} (a) we schematically depict a generic phenomenological
phase diagram of a system exhibiting a Gaussian quantum critical point (QCP). 
At $T=0$ the system can be tuned between an ordered and a disordered state
by varying a non-thermal control parameter $r$.  
At finite $T$, in the disordered phase, a crossover region, schematically represented here
with a line, separates the Fermi liquid and quantum critical regimes. 
%%%In conventional theory the crossover line is given by  $T_{\rm cr}\sim r^{z/2}$, 
%%%where $z$ is the dynamical exponent. 
For $r<0$, the $T_c$-line separates the phases of the system at finite $T$. The
classical critical region where non-Gaussian fluctuations occur is bounded by the
Ginzburg lines and vanishes as $T \to 0$. 

%The region of the phase diagram characterized by
%  strong classical fluctuations narrows down and vanishes at $T=0$.

By varying another system parameter the phase diagram can be continuously deformed so
that the transition becomes first order at sufficiently low $T$, below a
tricritical temperature $T^{\rm tri}$ (see Fig.~\ref{Phase_diag_pheno} (c)). For a
particular value of this parameter the tricritical point is located exactly at
$T=0$ so that the transition is second order for all $T>0$ but the scaling
properties of the system for $T\to 0$ are governed by the quantum tricritical point (QTCP)
rather than the quantum critical point. This special scenario displaying a 
different scaling behavior is depicted in Fig.~\ref{Phase_diag_pheno} (b).
The purpose of the paper is to compute how the crossovers between the
scenarios of Fig.~\ref{Phase_diag_pheno} (a) through (b) to
Fig.~\ref{Phase_diag_pheno} (c) occur. 
Even in scenario (a) a ``hidden'' tricritical point, which would be 
present at $T^{\rm tri}<0$, i.e. the vicinity to a first order transition,  
can affect the scaling behavior at higher temperatures.

%%%%%%%%%%%%%%%%%%%%%%%%%%%%%%%%%%%%%%%%%%%%%%%%%%%%%%%%%%%%%%%%%%%%%%%%%%%%%%%%%%%%%%%%

\section{Bare action}

The conventional description of quantum criticality, which we shall rely upon here, 
invokes the Hertz action.\cite{hertz76} This describes a bosonic mode overdamped by
particle-hole excitations across the Fermi level and is applicable under the assumption 
that the electronic degrees of freedom may be integrated out. 
The original framework of Hertz \cite{hertz76} and Millis \cite{millis93} was recently 
extended to account for a number of systems and phenomena not covered by the original 
studies. 
These include field-tuned quantum critical points, \cite{fischer_rosch05} metamagnetic 
transitions,\cite{millis_schofield02, SMGL02} phase transitions induced by a non-equilibrium 
drive \cite{mitra_takei06} as well as dimensional crossovers \cite{garst08} and quantum 
criticality involving multiple time scales. \cite{zacharias09} 
We focus on the case of discrete symmetry breaking, which can be described by a
scalar order parameter field $\phi$, and an action of the form
\begin{eqnarray}
 S[\phi] = 
 \frac{T}{2} \sum_{\omega_n} \int\frac{d^dp}{(2\pi)^d} \,
 \phi_p \left(Z_{\omega} \frac{|\omega_{n}|}{|\mathbf{p}|^{z-2}}
 + Z\mathbf{p}^{2} \right) \phi_{-p} + {\cal U}[\phi] \; .
 \label{eq:lagrangian}
\end{eqnarray}
Here $\phi_p$ with $p = (\mathbf{p},\omega_n)$ is the momentum representation of the
order parameter field, where $\omega_n = 2\pi n T$ with integer $n$ denotes the (bosonic) 
Matsubara frequencies. 
Momentum and energy units can be chosen such that the prefactors in front of 
$\frac{|\omega_{n}|}{|\mathbf{p}|^{z-2}}$ and $\mathbf{p}^{2}$ are equal to unity. 
The action is regularized in the ultraviolet by restricting momenta to 
$|\mathbf{p}| \leq \Lambda_0$. The term $\frac{|\omega_{n}|}{|\mathbf{p}|^{z-2}}$, 
where $z\ge 2$ is the dynamical exponent, effectively accounts for overdamping of 
the order-parameter fluctuations by fermionic excitations across the Fermi surface. 
The expression Eq.~(\ref{eq:lagrangian}) is valid for $|\mathbf{p}|$, and
$\frac{|\omega_{n}|}{|\mathbf{p}|^{z-2}}$ sufficiently small, which is the
limit relevant for the physical situation considered.\cite{millis93} 

The quantity ${\cal U}[\phi]$ is a local effective potential which we expand to 
sixth order in $\phi$:
\begin{equation}
 {\cal U}[\phi] =
 \int_0^{1/T}  d\tau \int \! d^d x \, U \left(\phi\left(x,\tau\right)\right) \; ,
\end{equation}
where
\begin{equation}
 U(\phi) = a_2 \phi^2 + a_4 \phi^4 + a_6 \phi^6  \; .
\label{eq:ef_potential}
\end{equation}
This ansatz assumes the absence of a field explicitly breaking the  
inversion symmetry.
We require $a_6 > 0$ to stabilize the system at large $|\phi|$. 
The bare effective potential yields the well known phase diagram, \cite{lawrie84} 
exhibiting a second-order transition for $a_4 > 0$, a first order transition for 
$a_4 < 0$ and a tricritical point at $a_2 = a_4 = 0$. Within the present model the 
mass parameter $a_2$ plays the role of the non-thermal control parameter ($r$) to
tune the transition. 
%%%We neglect the dependence of $a_2$ on temperature. For a discussion of the effects 
%%%of this dependence on quantum tricritical behavior see Ref.~\onlinecite{misawa08}.   
By varying the quartic coupling $a_4$ one can deform the phase diagram so that the 
transition is second or first order at low $T$ (compare Fig.~\ref{Phase_diag_pheno}). 
In conventional quantum criticality, the temperature dependence of the coefficients 
$a_2$, $a_4$, $a_6$ can be neglected, as it leads only to subleading corrections. 
However, it turns out that in a quantum tricritical regime the generic quadratic
temperature dependence of these coefficients yields a dominant contribution to the
temperature dependence of the correlation length.

%%%%%%%%%%%%%%%%%%%%%%%%%%%%%%%%%%%%%%%%%%%%%%%%%%%%%%%%%%%%%%%%%%%%%%%%%%%%%%%%%%%%%%%%

\section{Flow equations}
In the present study we apply the one-particle irreducible variant of the
functional RG. \cite{berges_review02, delamotte07, pawlowski_review07, gies_notes06} 
The derivation of the flow equations follows Ref.~\onlinecite{jakubczyk09}, where the present
quantum $\phi^6$ model was applied to analyze the effect of fluctuations on the
order of quantum phase transitions and the shapes of the finite $T$ phase
boundaries. The starting point is the exact functional evolution equation \cite{wetterich93}
\begin{eqnarray}
 \frac{\partial}{\partial \Lambda}\Gamma^{\Lambda}\left[\phi\right]=
 \frac{1}{2}\text{Tr}\frac{\partial_\Lambda R^{\Lambda}}{\Gamma^{(2)}\left[\phi\right]
 + R^{\Lambda}} 
\label{eq:flow_eqn}
\end{eqnarray}
describing the flow of the effective action $\Gamma^{\Lambda}[\phi]$, which is the
generating functional for one-particle irreducible vertex functions as the
infrared cutoff scale $\Lambda$ is reduced. 
Here $\Gamma^{(2)}\left[\phi\right] = \delta^{2}\Gamma^{\Lambda}[\phi]/\delta \phi^{2}$, 
$\text{Tr} = T \sum_{\omega_{n}} \int \frac{d^{d} p}{\left(2\pi\right)^{d}}$, 
finally $R^{\Lambda}$ denotes the cutoff function added to the inverse propagator 
to cut off modes with momentum below the scale $\Lambda$. 
We implement the Litim cutoff \cite{litim01} 
\begin{equation}
 R^{\Lambda}(\mathbf{p}) = 
 Z \left( \Lambda^{2}-\mathbf{p}^{2}\right)
 \theta\left(\Lambda^{2}-\mathbf{p}^{2} \right) \; .
\label{Litim_fun}
\end{equation}
For $\Lambda=\Lambda_0$ the quantity 
$\Gamma^{\Lambda}$ is just the bare action, while in the limit $\Lambda\to 0$
it converges to the full effective action, i.e., the Gibbs free energy. 
In most cases it is impossible to solve Eq.~(\ref{eq:flow_eqn}) in the full 
functional space. 
One may however cast it onto a suitably chosen set of flowing couplings. 
The approximation applied here amounts to assuming that the effective potential 
preserves the form given by Eq. (\ref{eq:ef_potential}) with flowing
couplings, while the inverse propagator retains its initial form
\begin{eqnarray}
 G^{-1}(\textbf{p},\omega_n) &=& 
 \Gamma^{(2)}\left[\phi=\phi_0\right] + R^{\Lambda}(\textbf{p}) \nonumber \\ 
 &=& Z_{\omega} \frac{|\omega_{n}|}{|\mathbf{p}|^{z-2}} + 
 Z\mathbf{p}^2 + 2a_2 + R^{\Lambda}(\textbf{p}) \; ,
\label{Green}
\end{eqnarray}
where $\phi_0$ is the minimum of $U(\phi)$.
The present study deals with situations where the quantum critical point is Gaussian. 
In the following we shall disregard the flow of $Z$, which is equivalent to neglecting 
the anomalous dimension of the order parameter field. 
Such a parameterization of the effective action does not capture the anomalous scaling 
behavior of the propagator in the narrow vicinity of the classical second-order phase 
transition at finite temperature. 
Analogous truncations retaining the flow of $Z$ were applied in 
Refs.~\onlinecite{jakubczyk09, jakubczyk08}. 
The results of Ref.~\onlinecite{jakubczyk08} indicate that universal aspects of the 
shape of the phase boundary are not affected by non-Gaussian thermal fluctuations.
In the following we shall also neglect the renormalization of the factor $Z_{\omega}$. 
The flow of $Z_{\omega}$ was computed in Ref.~\onlinecite{jakubczyk08}, and shown to 
be negligible. 
The present truncation reproduces the essential features of the system except for the narrow 
vicinity of the second order transition at $T>0$, where nonetheless the shape of 
the phase boundary is described correctly. 
Note, however, that the approach can be adapted to capture the anomalous 
dimension of the order parameter field \cite{jakubczyk08} and to deal with cases 
where the fixed point associated with the quantum critical point is not Gaussian.
\cite{strack09} 
The present truncation is analogous to the zeroth order derivative expansion,
\cite{berges_review02, delamotte07} which was applied extensively in the context 
of classical critical phenomena. The essential quantum ingredient present here is 
the Landau damping term $\frac{|\omega_n|}{|\textbf{p}|^{z-2}}$. 
On top of the derivative expansion we performed a polynomial expansion of the 
effective potential. 

\begin{figure}[ht!]
\begin{center}
\includegraphics[width=3.5in]{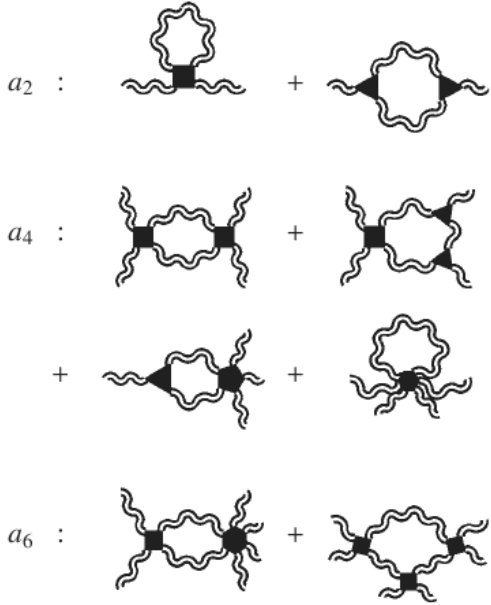}
\caption{Feynman diagrams representing the contributions to the
  flow equations for the couplings parameterizing the effective
  potential. Interactions involving an odd number of legs vanish 
  if the system is in the disordered phase.} 
\label{Diagrams}
\end{center}
\end{figure}

Relying on the truncation described above, the flow equation Eq. (\ref{eq:flow_eqn}) 
can be reduced to a set of three ordinary differential equations which determine the 
flow of the couplings $a_2$, $a_4$, $a_6$ as functions of the cutoff scale $\Lambda$. 
In practice, whenever the effective potential features non-zero minima at $\pm
\phi_0$, we find it more convenient to write the flow equations in terms of
the variables $\rho_0=\frac{1}{2} \phi_0^2$, $a_4$, $a_6$. The coupling $a_2$
is then obtained from  
\begin{equation}
 a_2=-4\rho_0(a_4+3a_6\rho_0)\;.
\label{a_2}
\end{equation}
As long as there exists a non-zero $\rho_0$, the evolution of the effective
potential is given by the flow equations derived in
Ref.~\onlinecite{jakubczyk09},
\begin{equation}
 \partial_t \rho_0 =
 2v_d Z^{-1}\Lambda^{d-2}\left(3+2\frac{6a_6\rho_0}{6a_6\rho_0+a_4}\right)l_1^d\;,
\label{flow_rho_0}
\end{equation}
\vspace{0.1cm}
\begin{equation}
 \partial_t a_4 =
 12v_d\Lambda^d\left[\frac{4}{3}l_2^d\frac{(30a_6\rho_0+3a_4)^2}{Z^2\Lambda^4}
 -5l_1^d\frac{a_6}{Z\Lambda^2}\right]-6\rho_0\partial_t a_6\;, 
\label{flow_a4}
\end{equation}
\begin{equation}
\partial_t a_6 =
 16v_d\Lambda^d\left[-\frac{8}{3}l_3^d\frac{(30a_6\rho_0+3a_4)^3}{Z^3\Lambda^6}+15l_2^d
 a_6\frac{30a_6\rho_0+3a_4}{Z^2\Lambda^4}\right], 
\label{flow_a6}
\end{equation}
where $v_d^{-1}=2^{d+1}\pi^{d/2}\Gamma(d/2)$ and $t = \log(\Lambda/\Lambda_0) \leq 0$.
The threshold \cite{berges_review02} functions $l_i^d(\delta)$ are determined from 
\begin{equation}
 l_0^d(\delta) =
 \frac{1}{4}v_d^{-1}\Lambda^{-d}\text{Tr} \, \frac{\partial_t R^{\Lambda}(\mathbf{p})}
 {Z_{\omega}\frac{|\omega_n|}{|\textbf{p}|} + Z\textbf{p}^2 + 
 R^\Lambda(\mathbf{p}) + Z\Lambda^2\delta}\; ,
\label{l0}
\end{equation}
\begin{equation}
 l_1^d(\delta) = -\frac{\partial}{\partial \delta}l_0^d(\delta)\; ,
\label{l1}
\end{equation}
\begin{equation}
 l_n^d(\delta) =
 -\frac{1}{n-1}\frac{\partial}{\partial \delta}l_{n-1}^d(\delta)\;,\;\;\; n\geq 2\;.
\end{equation}
In Eqs.~(\ref{flow_rho_0}, \ref{flow_a4}, \ref{flow_a6})
$\delta=\delta(\rho)=\frac{1}{Z\Lambda^2}(U'(\rho)+2\rho U''(\rho))$ and the
threshold functions are evaluated at the value of $\delta$ corresponding to 
$\rho_0$, that is, $l_n^d=l_n^d(\delta)|_{\rho=\rho_0}$. 
The flow equations are illustrated in terms of Feynman diagrams in Fig.~\ref{Diagrams}.

For $\rho_0=0$ the flow equations are simplified as the interaction vertices
involving an odd number of legs (see Fig.~\ref{Diagrams}) vanish. In terms of
the couplings $a_2$, $a_4$, $a_6$ they yield
\begin{equation}
 \partial_t a_2 = -24v_d\frac{a_4}{Z\Lambda^{2-d}}l_1^d\;,
\end{equation}
\begin{equation}
 \partial_t a_4 = 
 v_d\Lambda^d \left[\left(12\frac{a_4}{Z\Lambda^2}\right)^2l_2^d -
 60\frac{a_6}{Z\Lambda^2}l_1^d\right] \; ,
\end{equation}
\begin{equation}
 \partial_t a_6 = v_d \Lambda^d
 \left[-\frac{2}{3}\left(12\frac{a_4}{Z\Lambda^2}\right)^3l_3^d+
 \left(12\frac{a_4}{Z\Lambda^2}\right)\left(60\frac{a_6}{Z\Lambda^2}\right)l_2^d\right] 
 \;. 
\end{equation}
%

%%%\hspace{1cm}

%In the disordered phase (the global minimum of $U$ becomes zero) all the interactions 
%involving an odd number of legs vanish.
%  

%%%%%%%%%%%%%%%%%%%%%%%%%%%%%%%%%%%%%%%%%%%%%%%%%%%%%%%%%%%%%%%%%%%%%%%%%%%%%%%%%%%

\section{Linearized flow equations}

Essential features of the solutions to the equations provided in the previous
section (the values of the exponents describing the system in the quantum-critical 
regime in particular) can be computed analytically by taking only the dominant 
terms in the flow equations into account. 
Here we analyze the linearized version of the flow equations in spatial dimension $d>2$. 
%%%As we shall discuss in the following section, the case $d=2$ requires retaining also 
%%%terms quadratic in the interaction. 
We shall use the following rescaled variables:
\begin{equation}
 \delta = \frac{2a_2}{Z\Lambda^2},\;\; u = 
 \frac{a_4}{Z^2\Lambda^{4-d}},\;\; v = 
 \frac{a_6}{Z^3\Lambda^{6-2d}}\;.
\end{equation}
We linearize the flow equations around the zero temperature Gaussian fixed point
\begin{equation}
\partial_t\delta=-2\delta-48v_d l_1^d(\delta)u\;, 
\label{eq:delta_flow_i}
\end{equation}
\begin{equation}
\partial_t u =(d-4)u-60v_dl_1^d(\delta)v\;, 
\label{eq:u_flow_lin_i}
\end{equation}
\begin{equation}
 \partial_t v = (2d-6)v\;.
\label{eq:v_flow}
\end{equation}
All the resulting flow equations involve contributions from rescaling. In
addition, renormalization of the couplings $\delta$, $u$ by the diagrams
involving just one interaction vertex is retained. After this step, the terms
involved are analogous to those analyzed by Millis, \cite{millis93} where
$v=0$, so that only mass renormalization from the tadpole diagram was 
considered. Note that in the present case of discrete symmetry-breaking, 
Eq.~(\ref{eq:delta_flow_i}, \ref{eq:u_flow_lin_i}, \ref{eq:v_flow}) are applicable 
both in the symmetric and symmetry-broken phases.

In the above form we still find the flow equations rather hard to solve by
hand as the threshold function $l_1^d(\delta)$ involves integrals (see
Eqs.~(\ref{l0},\ref{l1})). Inspired by the seminal work by Millis,\cite{millis93} 
we additionally expand the function $l_1^d$ for 
$\tilde{T}=\frac{2\pi T Z_{\omega}}{Z\Lambda ^z}\ll 1$ and  $\tilde{T}\gg 1$. 
In the regime $\tilde{T}\ll 1$ the flow is dominated by quantum physics, 
while for $\tilde{T}\gg 1$ the quantum contributions to the flow equations 
are negligible compared to the classical ones. Similarly to
Ref.~\onlinecite{millis93} we shall integrate the flow equations from
$\Lambda=\Lambda_0$ to $\Lambda=\Lambda^*$ determined by $\tilde{T}(\Lambda^*)=1$
using the equations valid strictly speaking for $\tilde{T}\ll 1$. The result
will then serve as the initial condition for the solution of the flow from
$\Lambda^*$ to $\Lambda =0$, where we shall in turn use the asymptotic flow
equations valid in the classical sector $\tilde{T}\gg 1$. 

Performing the abovementioned expansion of the threshold function $l_1^d$ we find 
for $\tilde{T}\ll 1$
\begin{equation}
\partial_t \delta \approx -2\delta -48v_d\frac{4T}{\tilde{T}}\frac{u}{d+z-2}\;,
\label{eq:delta_flow_lin_ini}
\end{equation}
\begin{equation}
 \partial_t u \approx (d-4)u-60v_d\frac{4T}{\tilde{T}}\frac{v}{d+z-2}\; ;
\end{equation}
while for $\tilde{T}\gg 1$
\begin{equation}
 \partial_t \delta \approx -2\delta -48v_d\frac{2T}{d}u\;,
\label{eq:delta_flow_lin_fin}
\end{equation}
\begin{equation}
\partial_t u =(d-4) u -60 v_d \frac{2T}{d}v\;. 
\end{equation}
The above equations, together with Eq.~(\ref{eq:v_flow}), form a set of linear
ordinary differential equations of first order. Eq.~(\ref{eq:v_flow}) is
solved immediately. Plugging the result into the flow of $u$ and integrating,
we calculate $u(\Lambda)$ for $\Lambda>\Lambda^*$ and
$\Lambda<\Lambda^*$. Demanding that $a_4^r\equiv\lim_{\Lambda\to 0} 
a_4(\Lambda)=\lim_{\Lambda\to 0} Z^2\Lambda^{4-d}u(\Lambda)>0$ we find a condition for
the occurrence of a second order transition. If $a_4^r(T)=0$ at some $T\geq 0$,
there is a tricritical point 
at this $T$. We find that the transition is always second order at
sufficiently high $T$. It may turn first order as $T$ approaches zero. The
condition $a_4^r>0$ yields
\begin{equation}
 u_0 > -v_0 \left( C + C' T^{(d+z-2)/z}\right)\;,
\label{eq:condition_for_u}
\end{equation}
where $u_0 = u(\Lambda=\Lambda_0)$, $v_0 = v(\Lambda_0)$, and
$C$, $C'$ are positive constants (depending on $\Lambda_0$, $Z_\omega$, $Z$,
$d$ and $z$). The explicit expressions for these constants are given in the
Appendix. 
If the condition for $u_0$ given by Eq.~(\ref{eq:condition_for_u}) is fulfilled
for $T=0$, it clearly holds for any $T\geq 0$. In such case there is a line of
critical points terminating at $T=0$ with a QCP. This is the scenario (a) in
Fig.~\ref{Phase_diag_pheno}. For $v_0=0$ one
obviously recovers $u_0>0$ and the standard quantum critical behavior. A
quantum tricritical point exists if 
$u_0= u_0^{\rm tri}=-Cv_0$, which corresponds to the quantum
tricritical scenario (b) in Fig.~\ref{Phase_diag_pheno}.
In case the condition Eq.~(\ref{eq:condition_for_u})
is violated at $T=0$, it is still always fulfilled at sufficiently high
temperatures, namely for $T > T^{\rm tri}$, where
\begin{equation}
 T^{\rm tri} = [-(u_0/v_0+C)/C']^{z/(d+z-2)} \; .
\end{equation} 
In this case the transition becomes first order for $T<T^{\rm tri}$, 
which is scenario (c) in Fig.~\ref{Phase_diag_pheno}.  

We now discuss the flow of $\delta$. Plugging the solution obtained for
$u(\Lambda)$ into Eq.~(\ref{eq:delta_flow_lin_ini},
\ref{eq:delta_flow_lin_fin}), we integrate Eq.~(\ref{eq:delta_flow_lin_ini})
from $\Lambda_0$ to $\Lambda^*$. The solution at $\Lambda^*$ is the initial
condition for Eq.~(\ref{eq:delta_flow_lin_fin}). In cases where critical
behavior occurs, the correlation length can be extracted as
$\xi^{-2}\propto\lim_{\Lambda\to 0}\Lambda^2\delta$. 
The result has the following structure:
\begin{equation}
 \xi^{-2} = D_0(T) + D_1 T^{(d+z-2)/z} + v_0 D_2 T^{(2(d+z)-4)/z} \; ,
\label{eq:corr_length_lin}
\end{equation}
where $D_0$ depends on $\delta_0$, $u_0$ and $v_0$, while $D_1$ and $D_2$ do not
depend on $\delta_0$.
Notice that $D_1 = \bar D_1(u_0 - u_0^{\rm tri})$, where $\bar D_1 > 0$. 
Explicit expressions are quoted in the Appendix. 
The generically quadratic temperature dependence of the bare model parameters
$\delta_0$, $u_0$ and $v_0$ for low $T$ induces a corresponding quadratic 
temperature dependence of $D_0$, $D_1$, $D_2$. While the temperature dependence 
of $D_1$ and $D_2$ is always negligible, the quadratic $T$-dependence of 
$D_0(T) = D_0 + D'_0 T^2$ dominates in the tricritical regime (see below) and must 
therefore be kept.\cite{misawa08,misawa09}

Criticality occurs at $T=0$ when $\delta_0 = \delta_0^{\rm cr}$ such that $D_0 = 0$. 
At finite $T$ the behavior of $\xi^{-2}$ is governed by several terms.
The quantum critical term $D_1 T^{(d+z-2)/z}$ dominates for sufficiently
low $T$, provided $D_1 > 0$, and reproduces the result for $\xi^{-2}(T)$ derived 
already by Millis.\cite{millis93} Compared to that term, the quadratic temperature
dependence from $D_0(T)$ and the term $v_0 D_2 T^{(2(d+z)-4)/z}$ coming from the
$\phi^6$ interaction are subleading corrections.
On the other hand, if $u_0 = u_0^{\rm tri}$, the coefficient $D_1$ becomes zero and 
the correlation length exhibits a different behavior,
\begin{equation}
 \xi^{-2} = D'_0 T^2 + v_0 D_2 T^{(2(d+z)-4)/z} \; .
\end{equation}
The exponent of the second term is $3$ for $z=2$, $d=3$, and $8/3$ for $z=3$, $d=3$.
%%%These exponents can also be extracted from the numerical solution of the full
%%%flow equations, presented in Sec.~VI.
The latter exponent was obtained already from a self-consistent fluctuation
resummation.\cite{green05}
At low temperatures the ''trivial'' quadratic term thus dominates the temperature
dependence of $\xi$ in the tricritical regime.
The relevance of the quadratic temperature dependences of the bare parameters in the 
tricritical regime was noticed already by Misawa et al.\cite{misawa08,misawa09}
However, the exponent for the temperature dependence of the order parameter 
susceptibility obtained by these authors for the quantum tricritical regime 
turned out to be identical to that for conventional quantum criticality, that is, 
$3/2$ for $z=2$. 
That result, obtained from a self-consistent summation of staggered and homogeneous 
fluctuations, is clearly at variance with our result.\cite{fn1}

For quantum critical systems close to quantum tricriticality (small $D_1$), the
correlation length follows a tricritical temperature dependence above a crossover
temperature
\begin{equation}
 T_{\rm cross} = \left( \frac{D_1}{D'_0} \right)^{z/(2+z-d)}\;.
\label{Tcross1}
\end{equation}
For the special case of a very small $D'_0$, tricritical behavior with an
exponent $\frac{2(d+z)-4}{z}$ is observed for temperatures above
\begin{equation}
 T_{\rm cross} = \left( \frac{D_1}{v_0 D_2} \right)^{z/(d+z-2)} \; .
\label{Tcross2}
\end{equation}

Eq.~(\ref{eq:corr_length_lin}) also yields the shape of the
phase boundary. From the condition $\xi^{-2}=0$ one finds 
$T_c \sim |\delta_0-\delta_0^{\rm cr}|^{\psi}$, where 
$\psi = \psi^{\rm qc} = \frac{z}{d+z-2}$
for quantum criticality, reproducing the result by Millis.\cite{millis93}
This value crosses over to $\psi = \psi^{\rm tri} = \frac{1}{2}$ when the 
quartic coupling approaches the tricritical value $u_0^{\rm tri}$.
Only for the special case $D'_0 = 0$ one obtains
$\psi^{\rm tri} = \frac{z}{2(d+z)-4}$. For small $D'_0 \neq 0$ a crossover
between different exponents occurs.

As already stated, for $u_0 < u_0^{\rm tri}$ a classical tricritical point exists at
$T = T^{\rm tri}>0$. In Eq.~(\ref{eq:corr_length_lin}) this manifests itself by a
negative value of $D_1$. From the condition $\xi^{-2} = 0$ we find the shape of
the phase boundary above $T = T^{\rm tri}$ to follow
$|\delta_0-\delta_0^{\rm tri}| \approx a|T-T^{\rm tri}|+b|T-T^{\rm tri}|^2$. 
This reproduces a MFT result for classical tricritical points.\cite{lawrie84}  

From the solution for $\delta(\Lambda)$ we also read off the shape of the
crossover line separating the Fermi liquid and the quantum critical regimes. 
The condition for the occurrence of the quantum disordered (Fermi liquid) 
regime \cite{millis93} is that the correlation length becomes of the order 
of the inverse upper cutoff before classical scaling is reached,
that is, $\delta (\Lambda^*) > \Lambda_0^2$. From this condition we
find the shape of the crossover line as $\delta_0-\delta_0^{\rm cr}\sim
T^{2/z}$, as can also be deduced by a phenomenological
reasoning.\cite{sachdev_book, vojta_review03} 

The linearized flow equations discussed above are not applicable in two 
dimensions. The reason is easily understood from the diagrammatic interpretation 
provided in Fig.~\ref{Diagrams}. 
The tadpole diagram renormalizing the $a_2$ coefficient exhibits a logarithmic 
infrared singularity for $\xi \to \infty$ at $T>0$. 
The divergence is cured when one accounts for the renormalization of the quartic
interaction coupling via the terms of the order $a_4^2$. 
%%%This complication is avoided in the scheme applied in Ref.~\onlinecite{millis93}, 
%%%where the flow is terminated once $\delta$ is scaled out to sufficiently large
%%%values. 

%%%%%%%%%%%%%%%%%%%%%%%%%%%%%%%%%%%%%%%%%%%%%%%%%%%%%%%%%%%%%%%%%%%%%%%%%%%%%%%%%%%%%%%

\section{Numerical solutions to the flow equations}

Here we present results from direct numerical solutions to the flow equations
delivered in Sec. IV. Information concerning the physical state of the system
is read off from the renormalized values of the quantities $a_2$, $a_4$, $a_6$
in the limit $\Lambda \to 0$, $a_n^r=\lim_{\Lambda\to 0}a_n(\Lambda)$. In all
numerical computations we put $Z=1$, $Z_\omega =1$, $\Lambda_0=1$ and 
$a^0_6=a_6(\Lambda_0)=1$. We omit the
quadratic $T$-dependence of $a_2$ in this section. We analyze solutions to the
flow equations upon varying the parameters 
$a_2^0=a_2(\Lambda_0)$, $a^0_4=a_4(\Lambda_0)$ and $T$. 
We refer to Ref.~\onlinecite{jakubczyk09} for example plots of the
numerically computed phase boundaries including cases involving a first order
transition. Here we focus on the situation where $a^0_4$ is close to a
$a_4^{0,\rm tri}$, a value where a tricritical point exists at $T=0$ for some
$a^0_2$.  In the analytical calculation based on the linearized equations one finds
the quantum tricritical point for the choice 
$a_4^0 = a_4^{0,\rm tri} = -\frac{120\Lambda_0^{d+z-2}v_d}{\pi(d+z-2)^2}a_6^0$.
For $d=z=3$, $\Lambda_0 = a_6^0 = 1$ this gives 
$a_4^{0,\rm tri} \approx -0.0302358$. 
In the numerical solution for the full flow equations
a value for $a_4^{0,\rm tri}$ emerges on fine tuning $a^0_2$ and $a^0_4$,
such that $a^r_2\approx a^r_4\approx 0$, where the value for $a^0_4$ tends to
come out a bit (by less than 1\%) smaller than the analytical estimate.  

Once the transition line is established, we can fix $a^0_4$ and
investigate the quantum critical regime spanning over the quantum critical or
tricritical point. 
In particular, we extract the correlation length $\xi$ as function
of $T$. In the region of interest the system is typically ordered within the
bare model ($a^0_4<0$, $a^0_2>0$, but $(a^0_4)^2>4a^0_2a_6^0$), but becomes
disordered in the course of the flow, as the order parameter becomes zero for
some finite $\Lambda$. At this scale we have to switch to the set of
equations valid in the symmetric phase (see Sec.~IV). 

\begin{figure}[ht!]
\begin{center}
\includegraphics[width=3.0in]{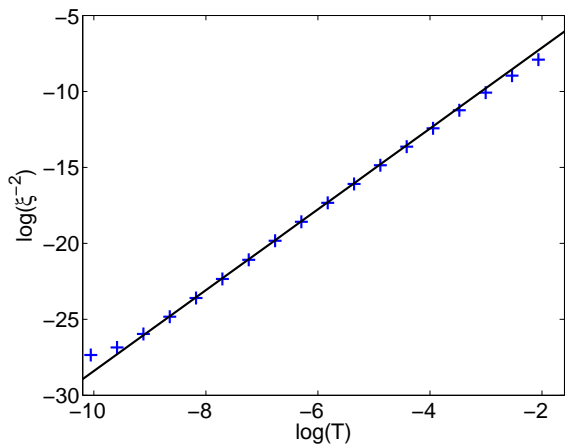}
\caption{(Color online) The logarithm of the inverse correlation length
  $\xi^{-2}$ versus $\log(T)$ in the quantum tricritical regime in
  $d=z=3$. The special choice of parameters is $a_4^0 = -0.030270317439$ and
  $a_2^0=0.00018421858$.}
\label{xi_tricrit_d3z3}
\end{center}
\end{figure}

We first show results for pure tricritical scaling of the inverse correlation
length $\xi^{-2}$ as a function of temperature (see Fig.~\ref{xi_tricrit_d3z3}).
We set $d=z=3$ and we choose the parameter $a^0_4\approx
a_4^{0,\rm tri}$ and $a^0_2$ tuned to the zero temperature quantum tricritical
point. This corresponds to the situation depicted in Fig.~1 (b), where the QTCP 
is approached from above.
Over a wide range of temperature we find that the slope $\epsilon=2.66$ gives 
a good fit to the numerical data, corresponding to the analytically
derived value $1/\psi^{\rm tri}= 8/3$. The leveling off at low temperatures
could be avoided by further fine-tuning $a^0_4$, while for higher temperature the
slope decreases as non-universal features start to play a role towards the cut-off 
scale. 

\begin{figure}[ht!]
\begin{center}
\includegraphics[width=3.0in]{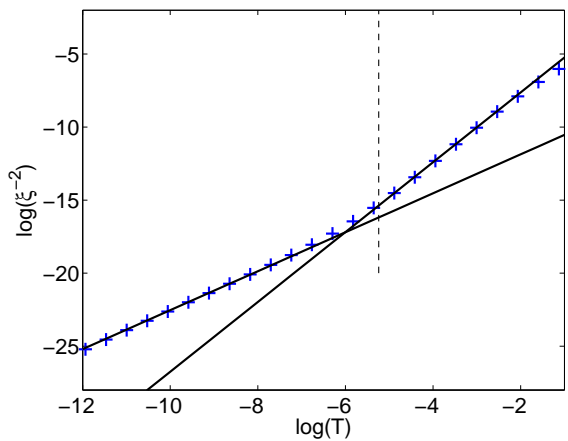}
\caption{(Color online) Correlation length versus temperature in the quantum
  critical regime in $d=z=3$. The plot parameters are $a_4^0 = -0.03$ slightly
  above the tricritical value and  $a_2^0=0.0001809109296$ tuned to the zero
  temperature transition point. The vertical dashed line gives the analytical 
  crossover temperature as obtained from Eq. (\ref{Tcross2}).  
  The fits (straight lines) are discussed in the text.} 
\label{xi_crossover_d3z3}
\end{center}
\end{figure}

If the value of $a^0_4$ is increased slightly above $a_4^{0,\rm tri}$ and
$a^0_2$ again tuned to the zero temperature quantum critical point the
crossover behavior in the correlation length as derived in equation
(\ref{eq:corr_length_lin}) appears.
Results of the numerical computations demonstrating this crossover behavior of
the correlation length are displayed in Fig.~\ref{xi_crossover_d3z3}. 

At low temperature the slope $\epsilon = 1.33$ fits well to the quantum critical
scaling $1/\psi_{\rm qc} = 4/3$. For larger temperatures the fit gives
$\epsilon = 2.39$, which lies about 10$\%$ below the value expected 
$1/\psi^{\rm tri}$. 
This can be explained with deviations from the power law at higher energy 
as noted in the discussion of Fig.~\ref{xi_tricrit_d3z3}. 
On choosing a value of $a_4^0$ closer to the tricritical one the crossover 
temperature decreases. The (low $T$) region characterized by the standard quantum 
critical scaling shrinks and tricritical scaling down to $T=0$ sets in, which is 
accompanied by $\epsilon$ approaching the value $8/3$ (see Fig.~3).

\begin{figure}[ht!]
\begin{center}
\includegraphics[width=3.0in]{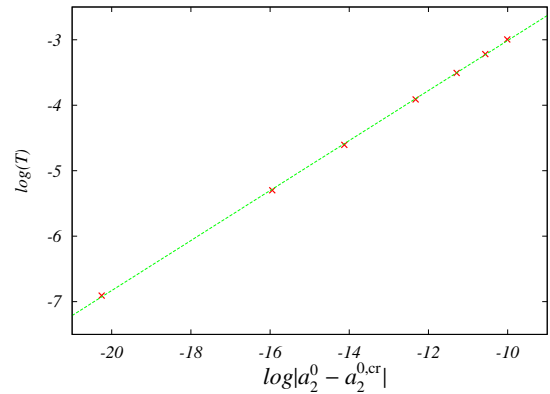}
\caption{(Color online) Transition line in $d=z=3$. The quartic coupling is
  chosen close to the tricritical value  $a_4^0=-0.03027 \approx a_4^{0,\rm tri}$.}
%Numerical fit of $\psi$ yields
%  $\psi\approx 0.378$, which can be compared to the value $3/8$ obtained
%  analytically in Sec.~V. When the quartic coupling is increased, the system
%  crosses over to the Hertz-Millis scaling with $\Psi^{\rm qc}=3/4$.
\label{phase_boundary_d3z3}
\end{center}
\end{figure}

We now turn to the phase boundary. The result computed for the quartic
coupling $a^0_4$ fine-tuned to its tricritical value is exhibited in
Fig.~\ref{phase_boundary_d3z3}. 
The shape of the phase boundary follows the power law $T\sim |a_2^0-a_2^{0,\rm cr}|^\psi$.
In the double logarithmic plot we find the slope $\epsilon\approx 0.378$, which compares 
well to the value $\psi^{\rm tri}=3/8$ obtained analytically in Sec.~V. When the 
quartic coupling is increased, the system shows the Hertz-Millis
scaling with $\psi^{\rm qc}=3/4$ for low temperature, and at larger
temperature tricritical scaling with $\psi^{\rm tri}=3/8$ is still visible for
the right choice of parameters (compare Ref.~\onlinecite{jakubczyk09}).

%%%%%%%%%%%%%%%%%%%%%%%%%%%%%%%%%%%%%%%%%%%%%%%%%%%%%%%%%%%%%%%%%%%%%%%%%%%%

\section{Summary}

We have presented a study of crossover behavior occurring in the vicinity of 
metallic quantum critical and tricritical points. The analysis is based on
renormalization group flow equations derived within the functional RG framework
applied to the Hertz action, retaining a $\phi^6$ term. It complements and
extends an earlier work (Ref.~\onlinecite{jakubczyk09}), providing
results in the quantum critical regime and delivering analytical insights. 
We focused on crossovers occurring in the temperature dependence of the
correlation length and the shape of the phase boundary in the quantum critical 
regime above the quantum critical and tricritical points in the phase diagram. 
The linearized form of the flow equations could be solved analytically in $d>2$,
yielding exponents for the power-laws obeyed by the correlation length and
phase boundary.
The analytical solutions were confirmed and complemented by a numerical evaluation of the 
full integro-differential flow equations.
In the quantum tricritical regime the power-law contribution to the inverse
correlation length generated by the $\phi^6$ interaction provides only a 
subleading correction to the generic quadratic temperature dependence induced
by the temperature dependence of the coefficients in the bare action.
For quantum critical systems close to quantum tricriticality, we computed the 
crossover temperature above which tricritical scaling is observed. 
In the situation where a tricritical point occurs at $T>0$, we obtained
an expression for the tricritical temperature, and recovered the shift exponent 
corresponding to the classical theory of tricriticality. 

%%%%%%%%%%%%%%%%%%%%%%%%%%%%%%%%%%%%%%%%%%%%%%%%%%%%%%%%%%%%%%%%%%%%%%%%%%%%%%%%

\begin{acknowledgments}
The authors would like to thank M. Nieszporski, S. Takei, H. Yamase for
valuable discussions, and S. Takei for a careful reading of the manuscript.
PJ was partially supported by  the German Science Foundation through the
research group FOR 723.  
\end{acknowledgments}

%%%%%%%%%%%%%%%%%%%%%%%%%%%%%%%%%%%%%%%%%%%%%%%%%%%%%%%%%%%%%%%%%%%%%%%%%%%%%%%%
 
\begin{appendix}

\section{Explicit expressions for the constants in Section V}

In this Appendix we give explicit expressions for the constants appearing in
the solutions to the flow equations in Sec. V. 
For the equations for $u(\Lambda)$ we have introduced
\begin{equation}
C=A\Lambda_0^z\;,
\end{equation}
and
\begin{equation}
C'=\Lambda_0^{2-d}\left(\frac{2\pi
    Z_{\omega}}{Z}\right)^{\frac{d-2}{z}}\left(B-\frac{2\pi Z_{\omega} }{Z}A\right)\;.
\end{equation}
For $\delta(\Lambda)$ one has 
\begin{equation}
D_0=\delta_0\Lambda_0^2+ A'\left(u_0+\frac12
  v_0 A\Lambda_0^{z}\right)\Lambda_0^{z+2}  \; ,
\end{equation}
and
\begin{eqnarray*}
D_1&=&\Lambda_0^{4-d}\left(u_0+ v_0A\Lambda_0^{z}\right)\left(\frac{2\pi
    Z_\omega}{Z}\right)^{\frac{d-2}{z}}\times  \\
&&
\times\left[B'-A'\left(\frac{2\pi Z_\omega}{Z}\right)\right]\;,
\end{eqnarray*}
and 
\begin{eqnarray*}
D_2&=& \frac{A'A\Lambda_0^{6-2d}}{2}\left(\frac{2\pi
    Z_\omega}{Z}\right)^{\frac{2(d+z)-4}{z}}+    \\
&&+B' \Lambda_0^{6-2d}\Bigg[\frac{B}{2}\left(\frac{2\pi Z_\omega}{Z}\right)^{\frac{2d-4}{z}}-
A\left(\frac{2\pi Z_\omega}{Z}\right)^{\frac{2d+z-4}{z}}\Bigg]
\; ,
\end{eqnarray*}

where 
\begin{equation}
A=120v_d\frac{Z}{\pi Z_\omega}\frac{1}{(d+z-2)^2}\;,
\end{equation}
$A'=96\,A/120$
%\begin{equation}
%A'=96v_d\frac{Z}{\pi Z_{\omega}}\frac{1}{(d+z-2)^2}\;,
%\end{equation}
and
\begin{equation}
 B=\frac{120v_d}{d(d-2)},
\qquad 
B'=\frac{96v_d}{d(d-2)}
\;.
\end{equation}

%If $Z$-factors are 1 one could define $a_4^0=\Lambda_0^{4-d} u_0$,
%$a_6^0=\Lambda_0^{6-2d} v_0$ and $a_4^{0,\rm
%  tri}=-\Lambda_0^{4-d}C\Lambda_0^zv_0=-\Lambda_0^{6-2d}v_0A\Lambda_0^{d+z-2}$, 
%then the first term in $D_1$
%becomes $(a_4^0-a_4^{0,\rm tri})$. $D_0$ and $D_2$ could also be simplified a bit.

\end{appendix}

%%%%%%%%%%%%%%%%%%%%%%%%%%%%%%%%%%%%%%%%%%%%%%%%%%%%%%%%%%%%%%%%%%%%%%%%%%%%%%%%%%%%%


\begin{thebibliography}{99}
\bibitem{fn1} Note that in the absence of an anomalous dimension the order parameter
 susceptibility and $\xi^2$ obey the same power-law.
\end{thebibliography}
\end{document}